\documentclass{article}

\usepackage{euscript}

\usepackage{latexsym}
\usepackage{amsmath, amssymb,graphics}
\usepackage[all]{xy}

\usepackage{fullpage}

\newtheorem{theorem}{Theorem}

\renewcommand{\(}{\begin{equation*}}
\renewcommand{\)}{\end{equation*}}
\newcommand{\bea}{\begin{eqnarray*}}
\newcommand{\eea}{\end{eqnarray*}}
\newcommand{\R}{{\mathbb R}}
\newcommand{\C}{{\mathbb C}}
\newcommand{\Z}{{\mathbb Z}}
\newcommand{\Q}{{\mathbb Q}}

\newcommand{\cF}{\ensuremath{\mathcal F}}

\newcommand{\cQ}{\ensuremath{\mathcal Q}}

\newcommand{\cM}{\ensuremath{\mathcal M}}

\newcommand{\cW}{\ensuremath{\mathcal W}}

\newcommand{\bo}{\raise-1mm\hbox{\Large$\Box$}}              
\newcommand{\g}{{\gamma}}

\newcommand{\G}{\Gamma}

\newcommand{\beq}{\begin{equation}}
\newcommand{\eeq}{\end{equation}}

\numberwithin{equation}{section}

\renewcommand{\(}{\begin{equation}}
\renewcommand{\)}{\end{equation}}

\newcommand{\CC}{{\mathbb C}}

\newcommand{\CP}{\CC \text{P}}

\def\R{{\mathbb R}}
\def\Z{{\mathbb Z}}
\def\Q{{\mathbb Q}}
\def\C{{\mathbb C}}

\def\1{{\bf 1}}

\def\<{\langle}
\def\>{\rangle}

\numberwithin{equation}{section}

\renewcommand{\(}{\begin{equation}}
\renewcommand{\)}{\end{equation}}

\begin{document}

%


\vspace{2em}
\def\thefootnote{\fnsymbol{footnote}}

\begin{center}
{\Large\bf 
Constraints on heterotic M-theory from s-cobordism }
\end{center}
\vspace{1em}

\begin{center}
\large Hisham Sati 
\footnote{e-mail: {\tt
hsati@math.umd.edu}\\
Current address: Department of Mathematics, University of Pittsburgh, 
139 University Place, Pittsburgh, PA 15260.
}
\end{center}

\begin{center}
Department of Mathematics\\
University of Maryland\\
College Park, MD 20742 
\end{center}

\vspace{0em}
\begin{abstract}
\noindent
 We interpret heterotic M-theory in terms of $h$-cobordism, that is 
 the eleven-manifold is a product of the ten-manifold times an interval 
 is translated into a statement that the former is a cobordism of the 
 latter which is a homtopy equivalence. In the non-simply connected 
 case, which is important for model building, the interpretation is then 
 in terms of $s$-cobordism, so that the cobordism is a simple-homotopy 
 equivalence. This gives constraints on the possible cobordisms depending on the
  fundamental groups
 and hence provides a characterization of possible compactification 
 manifolds using the Whitehead group-- a quotient of algebraic K-theory
 of the integral group ring of the fundamental group--
  and a distinguished element, the Whitehead torsion. 
 We also consider the effect on the dynamics via diffeomorphisms 
 and general dimensional reduction, 
 and comment on the effect on F-theory compactifications.

\end{abstract}


\tableofcontents

\section{Introduction}

A major goal of string theory is to provide a unification of fundamental interactions.
This includes constructing the standard model via string compactifications \cite{GSW2},
most notably via heterotic M-theory \cite{HW1} \cite{HW2}. 
Eleven-dimensional spacetime is taken to be an interval $I$ times 
a ten-manifold $M^{10}$, and with the two boundaries each supporting an $E_8$
gauge theory. One of the boundaries is called the hidden sector and the 
other is the visible sector, in which the structure group is broken down to 
a realistic symmetry group.
The ten-manifold $M^{10}$ 
is typically taken to be Minkowski space
$\R^{1,3}$ times a Calabi-Yau threefold $X^6$. 
In the visible sector one usually works with SU(5) or SO(10) $\subset E_8$
and breaks this group further (at least in principle) to the standard model
group (ideally) SU(3)$\times$SU(2)$\times$U(1). 
Physical and mathematical constraints on the (Calabi-Yau) manifold $X^6$
and bundles on $X^6$ are recently reviewed in \cite{He}.

\vspace{3mm}
Wilson lines are needed to break the gauge group from the grand unified (GUT) group
to the standard model group \cite{BOS} \cite{W85}. 
In order to introduce Wilson lines, the manifolds $M^{10}$ must have a 
nontrivial fundamental group. Starting with a simply connected Calabi-Yau
manifold, one gets a smooth non-simply connected Calabi-Yau manifold 
by dividing by a freely acting discrete symmetry $X^6 \mapsto X^6/\Gamma$, where
$\Gamma$ is a discrete group of finite order $|\Gamma|$, and the resulting
fundamental group is $\pi_1(X^6/\Gamma) =\Gamma$. Important choices for 
the finite group include $\Gamma=\Z_2$, which breaks SU(5) down to 
SU(3)$\times$SU(2)$\times$U(1), and $\Gamma= \Z_3\times \Z_3$ or $\Z_6$,
which break SO(10) down to SU(3)$\times$SU(2)$\times {\rm U(1)}^2$
\cite{GSW2}.

\vspace{3mm}
A major area of research involves choosing $\Gamma$ so that one gets the standard model,
not just as far as the symmetry groups are concerned but also accounting for example for correct
generations and spectra of particles. A sampler of fundamental groups of Calabi-Yau threefolds $X^6$ 
applied in the heterotic setting include:  $\Z_2$ \cite{BD},  $\Z_2 \times \Z_2$ 
 \cite{DOPR}, $\Z_3 \times \Z_3$ \cite{BOPR}  \cite{BHOP}, $\Z_8 \times \Z_8$ constructed in
 \cite{GP} on which 
rank 5 bundles are constructed in \cite{BBD}, abelian surface fibrations over $\C P^1$ with 
(abelianization of) fundamental group
$\Z_n \times \Z_n$ are considered in \cite{Sch} \cite{DGS}, complete intersection Calabi-Yau 
manifolds with fundamental groups which include \cite{CD}
$\Z_3$, $\Z_3 \times \Z_2$, $\Z_3 \times \Z_3$, $\Z_5$, 
$\Z_5 \times \Z_2$, $\Z_5 \times \Z_5$, and the quaternion group
$Q_8$, the latter being closely related to construction in \cite{BH}
 of Calabi-Yau threefolds with nonabelian 
fundamental groups, roughly speaking a semidirect product of $\Z_8$ with 
a quaternion group.
Torsion curves, important for instanton corrections to the heterotic 
minimally supersymmetric standard model (MSSM),
are studied in \cite{BKOS} for the quintic as well as for threefolds with fundamental groups
$\Z_3 \times \Z_3$.

\vspace{3mm}
Almost all known Calabi-Yau threefolds are simply connected. 
For example, only 16 out of about 500 million hypersurfaces in 
complex 4-dimensional toric varieties have nontrivial fundamental 
groups, and the only groups which occur are $\Z_2$, $\Z_3$ or 
$\Z_5$ \cite{BK}.
All elliptically fibered Calabi-Yau threefolds are simply connected, with the 
exception of fibrations over an Enriques base. In \cite{DOPW} elliptic 
fibrations without section, i.e. torus bundles, with 
nontrivial fundamental group are constructed. 
Another class of examples with no section is the Schoen family \cite{Sc}
which are fiber products of two rational elliptic surfaces.
Free finite group actions on these
are classified (under certain conditions) 
\cite{BD2} giving fundamental groups 
$\pi_1(X) \in \{ \Z_2, \Z_3, \Z_4, \Z_2\times\Z_2, \Z_5, \Z_6, \Z_2\times\Z_4, \Z_3\times\Z_3\ \}$.
\footnote{A free quotient of the manifold corresponding to $\Z_3 \times \Z_3$ by the quaternion
group is given in \cite{CD}.}
 In another class of threefolds, 
the complete intersections in products of projective spaces, an exhaustive search
\cite{Br} of the 7890 such threefolds leads to many interesting fundamental groups
including $\Z_i$ (for $i=2,3,4,5,6,8,10,12$), $\Z_2 \times \Z_j$
(for $j=2,4,8, 10$), $\Z_4 \times \Z_k$ (for $k=4,8$), $\Z_5 \times \Z_5$,
$\Z_8 \times \Z_8$,
as well as semidirect products $\Z_3 \rtimes \Z_4$,
$\Z_4 \rtimes \Z_4$, $\Z_5 \times \Z_{10}$
 and groups involving the quaternion group $Q_8$, namely
$\Z_2 \times Q_8$,  $\Z_2 \times \Z_2 \times Q_8$, 
$\Z_4 \rtimes Q_8$, $\Z_8 \rtimes Q_8$ (for a complete list see \cite{Br}).

\vspace{3mm}
In this paper we seek constraints on the possible fundamental groups 
coming from global considerations, namely from looking at the relation 
between the heterotic boundary and bounding M-theory. 
We first interpret this relation as a cobordism which connects
one boundary component to the other through the eleven-dimensional
bulk. 
We take one of the two boundary components 
and the bulk to be of the same homotopy type.
It is natural to ask when such cobordisms are trivial, that is when 
are they of product (or ``cylinder") form, as is usually the case in heterotic M-theory. 
When the fundamental groups of both the eleven-manifold
 $Y^{11}$ and the ten-manifold $M^{10}$ are trivial then we consider
 the cobordism as an $h$-cobordism ($h$ is for homotopy). When the fundamental groups
 are equal but nontrivial then we view heterotic M-theory as
 an $s$-cobordism ($s$ is for simple homotopy). 
 Since the dimension of the nontrivial part of $M^{10}$, namely
 the Calabi-Yau threefold, is six then the h-cobordism \cite{Mil0}
   and the s-cobordism \cite{Ker} \cite{Maz} \cite{Sta}
 theorems  can be applied. In both cases we are assuming that inclusions of 
the boundaries in $Y^{11}$ are homotopy equivalences. 
 The case when $\pi_1(Y^{11})$ is nontrivial is discussed extensively in 
 \cite{DMW-Spinc} in relation the partition functions and 
 to type IIA string theory. 

\vspace{3mm}
The obstruction to finding a cobordism that is of the cylinder type
is the Whitehead torsion of the inclusion $\tau (Y^{11}, M^{10})$, 
which is an element of the Whitehead group of the fundamental group
${\rm Wh}(\pi_1(M^{10}))$.
The Whitehead group is extensively studied and is well-known for finite groups (see \cite{Ol}),
which is the case we mainly study as such groups seem to be
the most interesting for model building. Given our identification of 
heterotic M-theory as an $s$-cobordism, we are able to 
identify fundamental groups that allow trivial cobordisms from the 
ones which do not. We view this as providing global consistency 
constraints on heterotic compactifications in view for model building. 

\vspace{3mm}
We summarize the main point of this article with 

\begin{theorem} 
Consider heterotic M-theory with the $E_8$ heterotic string theory on each of the two 
the boundary components. Then

\noindent $(i)$.  M-theory is an $s$-cobordism for one of the two connected components of the 
 boundary. 

\noindent $(ii)$. Consistency requires the Whitehead torsion, in the Whitehead group
of the integral group ring of the fundamental group of the boundary component, to vanish. 
\end{theorem}

\vspace{3mm}
Since the use of $h$- and $s$-cobordism and the Whitehead torsion is novel in the
context of heterotic M-theory and is perhaps not widely known in 
theoretical physics in general, we choose to take an expository route to arrive 
at our conclusions.  We provide the description of heterotic M-theory in terms of 
$h$ and $s$-cobordism in section \ref{sec cob}. Then in section \ref{sec ww}
we look at constraints on the fundamental group, coming from the Whitehead group
in section \ref{sec Wh} and from the Whitehead torsion in section \ref{sec tau}.
We provide many examples along the way and then in section \ref{sec ex}
we consider representative examples explicitly appearing in model building. 
We then consider the dynamical aspects in section \ref{sec dyn}, 
emphasizing the main points of this article. We first consider automorphisms,
including diffeomorphisms and issues of orientation, in section 
\ref{sec aut}, and then we consider dynamical aspects of compactifications
in section \ref{sec com}.

\section{Heterotic M-theory as an $h$-cobordism and $s$-cobordism}
\label{sec cob}


In this section we set up heterotic M-theory as a cobordism, first as an $h$-cobordism and 
then as $s$-cobordism. Viewed from M-theory the data involves an eleven-dimensional
manifold $Y^{11}$ which is a product $[0,1]\times M^{10}$ together with an $E_8$ bundle 
on each of $M^{10} \times \{0\}$ and $M^{10}\times \{1\}$. We will consider this from 
a ten-dimenional point of view, where we will have a cobordism taking one boundary 
component to the other.

\paragraph{$H$-cobordism.}
A compact connected eleven-manifold $Y^{11}$ whose boundary $\partial Y^{11}$
is the disjoint union of two closed manifolds $M^{10}$ and $M'^{10}$,
$\partial Y^{11}=M^{10} \cup M'^{10}$,
is called an $h$-cobordism, provided the inclusions of $M^{10}$ into 
$Y^{11}$ and of $M'^{10}$ into $Y^{11}$ are both homotopy equivalences. 
The pair $(Y^{11}, M^{10})$ is called a $h$-cobordism with base 
$M^{10}$ and top $M'^{10}$. 
A smooth $h$-cobordism is one where
$Y^{11}$ is a smooth manifold. 
A trivial or product $h$-cobordism is of the form $M^{10} \times [0,1]$. 
If $Y^{11}$ is simply-connected, then the $h$-cobordism theorem can be applied
(see \cite{Mil0}) to give that 
$Y^{11}$ is diffeomorphic to the product $M^{10}\times [0,1]$.
This is the configuration that is usually considered in heterotic M-theory \cite{HW1} \cite{HW2}. 

\vspace{3mm}
We can consider a more detailed description, which will be useful in section 
\ref{sec tau} and section \ref{sec dyn}.  
An eleven-dimensional cobordism 
$(Y^{11}; M_0^{10}, f_0, M_1^{10}, f_1)$ consists of a compact oriented eleven-manifold 
$Y^{11}$, two closed ten-manifolds $M_0^{10}$ and $M_1^{10}$, a disjoint 
decomposition $\partial Y^{11}=\partial_0 Y^{11} \coprod \partial_1Y^{11}$ of the boundary
$\partial Y^{11}$ of $Y^{11}$ and orientation-preserving 
diffeomorphisms $f_0: M_0^{10} \to \partial_0Y^{11}$ and $f_1: (M_1^{10})^- \to \partial_1Y^{11}$.
By $X^{-}$ we mean the manifold $X$ taken with the opposite orientation. On the 
boundary $\partial Y^{11}$ we use the orientation with respect to the decomposition 
$TY^{11}=T\partial Y^{11} \oplus \R$ coming from an inward normal field
to the boundary. 
If $\partial_0Y^{11}=M_0^{10}$, $\partial_1Y^{11}=(M_1^{10})^-$, 
and $f_0$ and $f_1$ are the identity maps, then the $h$-cobordism can be 
referred to as $(Y^{11};\partial_0Y^{11}, \partial_1Y^{11})$.
An $h$-cobordism over $M_0^{10}$ is trivial if it is diffeomorphic 
relative $M_0^{10}$ to the trivial $h$-cobordism 
$(M_0^{10}\times [0,1]; M_0^{10}\times \{0\}, (M_0^{10}\times \{1\})^{-})$.

\paragraph{The fundamental group.}
The $h$-cobordism theorem can be applied only when the fundamental group is
trivial.
Next we consider the more interesting case when the fundamental group 
is not necessarily trivial. We will assume that $\pi_1(Y^{11})\cong \pi_1(M^{10})$.
The fundamental group functor takes products to products, that is, the fundamental 
group is
 multiplicative. For $M^{10}=\R^{1,3}\times X$, we have 
 $
 \pi_1(M^{10})\cong \pi_1(\R^{1,3}) \times \pi_1(X) \cong \pi_1(X)$, 
so that the fundamental group of $M^{10}$ is determined by that of the 
Calabi-Yau threefold $X$. 
The generalization from Minkowski to other four-dimensional spacetimes 
gives an obvious modification, which depends on whether or not the latter 
is simply connected. 
Next we consider the appropriate 
description of heterotic M-theory when $\pi_1(M^{10})\neq 0$.
By our assumption, this is equivalent to taking $\pi_1(Y^{11}) \neq 0$, 
considered in \cite{DMW-Spinc}.

\paragraph{$S$-cobordism.}
 Let $M^{10}$ be a connected compact 10-manifold with fundamental group
$\G$, and consider the family $\cF$ of all $h$-cobordisms 
built on $M^{10}$. These are connected compact 11-manifolds $Y^{11}$ with 
exactly two boundary components, one of which is $M^{10}$ and the 
other of which is some other manifold $M'^{10}$ such that $Y^{11}$ is homotopy 
equivalent to both $M^{10}$ and $M'^{10}$. There is a map
$\tau : \mathcal{F} \to {\rm Wh}(\G)$ called the 
{\it Whitehead torsion} which induces a natural one-to-one correspondence 
from $\mathcal{F}/\sim$ to 
${\rm Wh}(\G)$, where $\sim$ is the equivalence relation induced by 
diffeomorphisms
$Y^{11} \to Y'^{11}$ which are the identity 
on $M^{10}$. If $Y^{11}$ is the ``trivial" $h$-cobordism 
$Y^{11}=M^{10} \times [0,1]$, then $\tau (Y^{11})=1$. 
This is an application of the Barden-Mazur-Stallings
theorem \cite{Ker} \cite{Maz} \cite{Sta} (see \cite{Ro} for a review).

\vspace{3mm}
If the fundamental group $\G$ is such its Whitehead group  ${\rm Wh}(\G)$ is trivial,
then certainly the Whitehead torsion will vanish and we are back to the 
case of an $h$-cobordism. Consequently, $Y^{11}$ is diffeomorphic (relative $M^{10}$) 
to a product $M^{10} \times [0,1]$. 
In particular, the other boundary component $M'^{10}$
is diffeomorphic to $M^{10}$. 
There is a bijection, given by the Whitehead torsion
$\tau (Y^{11}, M^{10})$, between the set of diffeomorphism classes of $h$-cobordisms
$(Y^{11},M^{10})$ with a given base $M^{10}$ and the set ${\rm Wh}(\pi_1(M^{10}))$. The cylinder 
corresponds to 0 under this bijection. We will consider this in much more detail in the 
following sections.

\vspace{3mm}
Note that there are several versions of the the $s$-cobordism (and $h$-cobordism)
theorem
depending on the category of spaces within which we are working; for example we
could work with homeomorphisms rather than diffeomorphisms (but here we are
assuming all spaces to be smooth).
However, if we start with a homotopy equivalence then we might not
be able to extend it to a diffeomorphism. 
Consider a large class of manifolds called {\it aspherical}, which are ones
for which all homotopy groups vanish except the first one, i.e.
the fundamental group. Let $Y$ and $M$ be aspherical 
spaces and let $\alpha: \pi_1( M) \to \pi_1(Y)$ be an isomorphism. 
Then, by the Theorem of Hurewicz, $\alpha$ is induced by
a homotopy equivalence. It is an open conjecture of Borel from 1955 that this 
can be extended to a homeomorphism. 
The strengthening to 
smooth manifolds fails \cite{DH}.

\vspace{3mm}
Note that we can work in category of spaces other than that of smooth manifolds,
since the $h$- and $s$-cobordism arguments work for piecewise linear (PL) 
and topological spaces.
This implies, for example, that orbifolds are also included in our discussion, for 
which we would choose the category of topological spaces.

\section{The Whitehead group and Whitehead torsion}
\label{sec ww}

We now consider the Whitehead group and Whitehead torsion in our setting 
of heterotic M-theory
via algebraic K-theory of the group ring of the fundamental group and 
give the main properties which are useful for us.

\subsection{The Whitehead group}
\label{sec Wh}
Algebraic K-theory roughly characterizes how, in passing from a
field to an arbitrary ring, notions of linear algebra related to the general linear group and
vector spaces might extend. One measure of failure of such an extension is 
$K_1(R)$, the algebraic K-theory of an associative ring $R$. 
Let $\widetilde{K}_1(R)$ be the cokernel of the map $K_1(\Z) \to K_1(R)$ induced by
the canonical ring homomorphism $\Z \to R$. 
Since $\Z$ is a ring with Euclidean algorithm then 
the homomorphism det: $K_1(\Z) \to \{ \pm\}$, given by $[A] \mapsto {\rm det}(A)$, 
is a bijection. Hence $\widetilde{K}_1(R)$ is the quotient of $K_1(R)$
by a cyclic group of order two generated by the class of the $1\times1$-matrix  
$(-1)$. 
We are interested in the case when 
$R$ is a group algebra $\Z[\Gamma]$ of the fundamental group $\Gamma=\pi_1(M^{10})$, 
that is in 
integer linear combinations of elements of $\G$.  
Define the {\it Whitehead group} ${\rm Wh}(\Gamma)$ of a group $\Gamma$
to be the cokernel of the map $\Gamma \times \{\pm \} \to K_1(\Z[\Gamma])$ which
sends $(\g, \pm 1)$ to the class of the invertible $1\times1$-matrix $(\pm \g)$. 
In other words, 
${\rm Wh}(\G)$ is the quotient 
of $K_1(\Z[\G])$ by the image of $\{ \pm \g : \g \in \G\}$, that is
\(
{\rm Wh}(\G)= K_1(\Z[\G]) / \{ \pm \g : \g \in \G   \}\;.
\)
The zero element $0 \in {\rm Wh}(\G)$ is represented by the identity matrix
$I_n$ for any positive 
integer $n$.

\vspace{3mm}
Note that one can define the Whitehead group of the fundamental group by 
choosing a base point, as is usual in the fundamental group. However, the end
result will be independent of the choice of the base point. Therefore, one should
think of $\pi_1(M)$ in ${\rm Wh}(\pi_1(M))$ as the {\it fundamental groupoid}
of $M$. Note also that the Whitehead group can be viewed either additively or
multiplicatively. In the first point of view, this corresponds to adding two cobordisms
by connecting one `cylinder' to another over a ten-dimensional section, while 
an instance of the second point of view is a 'flip'.

\paragraph{Example 1. Trivial case.} Consider the case when the fundamental group is trivial.
Then the group algebra $\Z[1]=\Z$ is a ring with a Gaussian algorithm, so that the
determinant induces an isomorphism $K_1(\Z) \buildrel{\cong}\over{\to} \{\pm\}$ and
the Whitehead group ${\rm Wh}(\{1\})$ of the trivial group vanishes. Hence any $h$-cobordism 
over a simply-connected closed $M^{10}$ is trivial. 
Thus, as expected in this case, $s$-cobordism reduces to $h$-cobordism.

\paragraph{Example 2. Finite cyclic groups.}
 ${\rm Wh}(\G)$ is torsion-free for a finite cyclic group. 
For example,  ${\rm Wh}(\Z_p)$, $p$ odd prime, is the free abelian 
group of rank $(p-3)/2$ and ${\rm Wh}(\Z_2)=0$.  

\vspace{3mm}
We will consider many more examples in section \ref{sec ex}.

\paragraph{Properties of the Whitehead group.} We are interested in the case when 
the fundamental group $\Gamma$ is a finite group. For such a group the following useful
properties hold  \cite{Mil} \cite{ADS} \cite{Ol}
\begin{enumerate} 
\item {\it Functoriality:} ${\rm Wh}(\Gamma)$ is a covariant functor of $\Gamma$, that is, 
any homomorphism $f: \Gamma_1 \to \Gamma_2$ induces a homomorphism 
$f_*: {\rm Wh}(\Gamma_1) \to {\rm Wh}(\Gamma_2)$.

\item {\it Trivial group:} Let $\G=\pi_1(M)$ be trivial. Then from $K_1(\Z)=\Z_2$ one gets
${\rm Wh}(\pi_1(M)) = {\rm Wh}(1)=1$. This is example 1 above.

\item {\it Low rank:} Whitehead showed that ${\rm Wh}(\Gamma)=1$ if $|\Gamma| \leq 4$.
This implies, for instance, that $\Z_2$, $\Z_3$, $\Z_4$ and $\Z_2 \times \Z_2$ have 
trivial Whitehead group and hence lead to (desirable) trivial $h$-cobordisms.

\item {\it Rank:} By a result of Bass, ${\rm Wh}(\Gamma)$ is a
finitely generated abelian group of rank $r(\Gamma)-q(\Gamma)$, where 
$r(\Gamma)$ is the number of 
irreducible real representations of $\Gamma$ and $q(\Gamma)$ is the number of 
irreducible rational representations of $\Gamma$. 
Explicitly, $q(\Gamma)$ is the number of conjugate classes of cyclic subgroups of 
$\Gamma$ and $r(\Gamma)$ is the number of conjugate classes of unordered pairs 
$\{ \g, \g^{-1}\}$.

\item {\it Free product:} The Whitehead group of a free product 
is multiplicative, ${\rm Wh}(G \ast H)={\rm Wh}(G) \oplus {\rm Wh}(H)$.
Unfortunately, there is no corresponding formula for Cartesian products.
For example, ${\rm Wh}(\Z_3)=0$ and ${\rm Wh}(\Z_4)=0$ but 
${\rm Wh}(\Z_3 \times \Z_4) \cong \Z$. More on this will be discussed in section 
\ref{sec ex}.

\end{enumerate}

\paragraph{The torsion subgroup.}
We have seen above that the Whitehead group of a cyclic group $\Z_p$ of prime
order $p$ is torsion-free. While these groups form an important  class of 
fundamental groups we are considering, we should consider other cases as well. 
In particular, there could be groups $\G$ for which ${\rm Wh}(\G)$ is torsion. 
The torsion in the algebraic K-group is 
${\rm Tor}(K_1(\Z[\G]))=(\pm) \times \G^{\rm ab} \times SK_1(\Z[\G])$, where 
$\G^{\rm ab}$ is the abelianization of $\G$ (that is the first homology group $H_1(M^{10})$)
and 
$SK_1(\Z[\G])=\ker (K_1(\Z[\G]) \to K_1(\Q[\G]))$. This kernel of the 
change of coefficients homomorphism is the full torsion subgroup of
${\rm Wh}(\Z[\G])$.

%
%

%
%
%
%
%
%
%
%
%

%
%
%

\paragraph{Properties of the torsion subgroup.} The torsion subgroup $SK_1(\Z[\G])$ of ${\rm Wh}(\G)$ is highly nontrivial \cite{ADS} \cite{Wa74} \cite{Ol}. Some of the useful properties are

\begin{enumerate}
\item 
The torsion subgroup of 
${\rm Wh}(\Gamma)$ is isomorphic to $SK_1(\Z[\Gamma])$.

\item The torsion in ${\rm Wh}(\G)$ comes from $SL(2,\Z[\G])$. 

\item $SK_1(\Z[\Gamma])$ is non-vanishing for 
all groups of the form $\Gamma \cong (\Z_p)^n$, $n\geq 3$ and 
$p$ an odd prime.

\item $SK_1(\Z[\Gamma])=1$ if $\Gamma \cong \Z_{p^n}$ or 
$\Z_{p^n}\times \Z_p$ (for any prime $p$, and any $n$),
if $\Gamma\cong (\Z_2)^n$ (any $n$), or if $\Gamma$ is any
dihedral, quaternion, or semidihedral 2-group. 

\item The classes of finite groups $\Gamma$ for which ${\rm Wh}(\Gamma)=1$,
or $SK_1(\Z[\Gamma])=1$, are {\it not} closed under products. 
This provides many nontrivial examples using products.

\end{enumerate}

For a finitely generated fundamental group $\G$ the vanishing of the 
Whitehead group ${\rm Wh}(\G)$ is equivalent to the statement that each
$h$-cobordism over a closed connected $M^{10}$ is trivial. 
Knowing that all $h$-cobordisms over a given manifold are trivial is 
useful, but strong. Alternatively, we could have ${\rm Wh}(\G)$ nontrivial
yet the distinguished element, the Whitehead torsion $\tau$ is zero.

\subsection{Whitehead torsion}
\label{sec tau}
The Whitehead torsion, which is essentially a linking matrix for handles in the
handle decomposition of the manifold, serves as an obstruction to 
the reduction of an $h$-cobordism to a product. 
We have encountered above many situations where the 
Whitehead group is not trivial. In certain cases these elements,
including the distinguished element given by the Whitehead torsion,  
can be characterized. This characterization can be geometric due to 
the realization theorem which says that every Whitehead torsion 
comes from a manifold (see \cite{Ker}).

\vspace{3mm}
First, note that a map $f: Y^{11}\to M_0^{10}$ induces a homomorphism
$f_*: {\rm Wh}(\pi_1(Y^{11})) \to {\rm Wh}(\pi_1(M_0^{10}))$ on the 
corresponding Whitehead groups such that 
${\rm id}_*={\rm id}$, $(g \circ f)_*=g_* \circ f_*$, and 
$f \simeq g$ implies that $f_*=g_*$. 
Next, the Whitehead torsion of our eleven-dimensional $h$-cobordism 
$(Y^{11}; M_0^{10}, f_0, M_1^{10}, f_1)$ over $M_0^{10}$, 
\(
\tau (Y^{11}, M_0^{10}) \in {\rm Wh}(\pi_1(M_0^{10}))\;,
\)
is defined to be the preimage of the Whitehead torsion
$\tau (M_0^{10}\buildrel{f_0}\over{\longrightarrow} \partial_0Y^{11} 
\buildrel{\iota_0}\over{\longrightarrow} Y^{11})\in {\rm Wh}(\pi_1(Y^{11})$ under the 
isomorphism $(\iota_0 \circ f_0)_*:
{\rm Wh}(\pi_1(M_0^{10})) \buildrel{\cong}\over{\longrightarrow}
{\rm Wh}(\pi_1(Y^{11}))$, where 
$\iota_0: \partial_0Y^{11} \hookrightarrow Y^{11}$ is the inclusion
(see \cite{KL}). 
Next we will consider the simple situation when the diffeomorphisms
are the identity.

\paragraph{Geometric definition of Whitehead torsion.}
There is a description of Whitehead torsion at the level of 
chain complexes \cite{Mil} \cite{Co}.
Let $\cW(M^{10})$ be the collection of all pairs of finite complexes $(Y^{11}, M^{10})$ 
such that $M^{10}$ is a strong deformation retract of $Y^{11}$. For any two objects
$(Y^{11}_1, M^{10})$, $(Y^{11}_2, M^{10}) \in \cW$ define 
an equivalence $(Y^{11}_1, M^{10}) \sim (Y^{11}_2, M^{10})$ if and only if
$Y^{11}_1$ and $Y^{11}_2$ are simple homotopically equivalent relative to the 
subcomplex $M^{10}$. 
Define ${\rm Wh}(M^{10})= \cW/\sim$ and 
let $[Y^{11}_1, M^{10}]$ and $[Y^{11}_2, M^{10}]$ be two classes
in ${\rm Wh}(M^{10})$. For
$Y^{11}_1 \bigsqcup_{M^{10}} Y^{11}_2$, the disjoint union of $Y^{11}_1$ and 
$Y^{11}_2$ identified along the common subcomplex $M^{10}$, an abelian
group structure 
can be defined on the Whitehead group ${\rm Wh}(M^{10})$ by 
$[Y^{11}, M^{10}] \oplus [Y^{11}_2, M^{10}]=
[Y^{11}_1 \bigsqcup_{M^{10}} Y^{11}_2, M^{10}]$.

\vspace{3mm}
The universal cover
$(\widetilde{Y}^{11}, \widetilde{M}^{10})$
of an element $(Y^{11}, M^{10})$ in $\cW$
can be equipped with the CW-complex structure lifted from the 
CW-structure of $(Y^{11}, M^{10})$. The inclusion 
$\widetilde{M}^{10} \subset \widetilde{Y}^{11}$ is a homotopy equivalence. 
Let $C_*(\widetilde{Y}^{11}, \widetilde{M}^{10})$ be the cellular chain complex
of $(\widetilde{Y}^{11}, \widetilde{M}^{10})$.
The covering action of $\pi_1(Y^{11})$ on 
$(\widetilde{Y}^{11}, \widetilde{M}^{10})$ induces an action on 
$C_*(\widetilde{Y}^{11}, \widetilde{M}^{10})$ and makes it 
a finitely generated free acyclic chain complex of $\Z[\pi_1(Y^{11})]$-modules. 
In addition to the boundary map $\partial$, there is 
a contraction map $\delta$ of degree $+1$ on 
$C_*(\widetilde{Y}^{11}, \widetilde{M}^{10})$
such that $\partial \delta + \delta \partial ={\rm id}$ and $\delta^2=0$. 
The module homomorphism 
$\partial + \delta: \bigoplus_{i=0}^\infty C_{2i+1}(\widetilde{Y}^{11}, \widetilde{M}^{10})
\to
\bigoplus_{i=0}^\infty C_{2i}(\widetilde{Y}^{11}, \widetilde{M}^{10})
$
is an isomorphism of $\Z[\pi_1(Y^{11})]$-modules. 
The image and the range of this homomorphism are finitely generated free 
modules with a basis we choose coming from the CW-structure on 
$(\widetilde{Y}^{11}, \widetilde{M}^{10})$.
Consider the matrix of this homomorphism 
$\partial + \delta$ which is an invertible matrix with entries
in $\Z[\pi_1(Y^{11})]$ and hence lies in $GL(n, \Z[\pi_1(Y^{11})])$ for some $n$. 
Now take the image of this matrix in ${\rm Wh}(\pi_1(Y^{11}))$
via an isomorphism $\tau$, sending $(Y^{11}, M^{10})$ 
to ${\rm Wh}(\pi_1(Y^{11}))$.

\vspace{3mm}
More explicitly, let 
$
 \cdots 
 \buildrel{\partial}\over{\longrightarrow}
C_{i+1}
 \buildrel{\partial}\over{\longrightarrow}
C_{i}
 \buildrel{\partial}\over{\longrightarrow}
\cdots 
C_{0}
 \buildrel{\partial}\over{\longrightarrow}
0
$
be the complex which calculates the homology $H_*(Y^{11}, M^{10};\Z[\G])$ of the
inclusion $M^{10} \subset Y^{11}$.
Each $C_i$ is a finitely generated free $\Z[\G]$-module. 
Up to orientation and translation by an element in $\G$, each $C_i$ 
has a preferred basis over $\Z[\G]$ coming from the $i$-simplices 
added to get from $M^{10}$ to $Y^{11}$ in some triangulation of the
universal covering spaces. The group $Z_i$ of $i$-cycles is the kernel of 
$\partial : C_i \to C_{i-1}$ and the group $B_i$ of $i$-boundaries is the image
of $\partial: C_{i+1} \to C_i$. Since $M^{10} \subset Y^{11}$ is a deformation 
retract, homotopy invariance of homology gives that $H_*=0$, so that 
$B_*=Z_*$. 
Let $\cM_i \in GL(\Z[\pi_1(M^{10})])$ be the matrices representing the 
isomorphism $B_i \oplus B_{i-1} \cong C_i$ coming from 
a choice of section $0 \to B_i \to C_i \to B_{i-1} \to 0$.
Let $[\cM_i] \in {\rm Wh}(\pi_1(M^{10}))$ be the corresponding equivalence classes.
The Whitehead torsion is then 
\(
\tau(Y^{11}, M^{10})= \sum (-1)^i [\cM_i] \in {\rm Wh}(\pi_1(M^{10}))\;.
\)
Note that the Whitehead group is identified as a quotient of $K_1(\Z[\G])$ by the 
subgroup generated by the units of the form $\pm \g$ for $\g \in \G=\pi_1(M_0^{10})$. 
In the present context, this ensures the independence of the choice of 
$\Z[\G]$-basis within the cellular equivalence class of $\Z[\G]$-bases.

\paragraph{Properties of Whitehead torsion.} 
The Whitehead torsion has existence and uniqueness properties. 

\vspace{2mm}
\noindent 1. {\it Existence.} Given $\alpha \in {\rm Wh}(\pi_1(M^{10})$, there exists an
$h$-cobordism $Y^{11}$ with $\tau (Y^{11})=\alpha$. This implies that if the Whitehead group
is nontrivial then we can find a cobordism for every element in that group. In order to get
a trivial $h$-cobordism, that is one of cylinder type, we have to make sure that the 
element ${\rm Wh}(\G)$ we identify for our spaces will be the zero element. 
This is of course not guaranteed to occur.

\vspace{2mm}
\noindent 2. {\it Uniqueness.} $\tau (Y^{11})= \tau (Y'^{11})$ if and only if there exists
a diffeomorphism $f: Y^{11} \to Y'^{11}$ such that $f|_M={\rm id}_M$.
This tells us that we are allowed to ``deform" $Y^{11}$ in a nice way and still be able to 
get the same type of cobordism. In particular, for $Y^{11}$ with $\tau(Y^{11})=0$ we
can always find a diffeomorphic $Y'^{11}$ for which the property that the Whitehead torsion
is zero is preserved. 

\paragraph{Elements of finite order in the Whitehead group.}
We have seen that the Whitehead group of products of 
finite cyclic groups may contain torsion. Elements of finite order can be 
characterized as follows \cite{Mil}.
Consider an orthogonal representation $\G \to O(n)$ of the finite 
group $\G$. This representation gives rise to a ring homomorphism 
$\rho: \Z[\G] \to {\rm \mathbb{M}}_n (\R)$, where ${\rm \mathbb{M}}_n(\R)$ is the algebra of $n \times n$
matrices over the real numbers. This induces a group homomorphism 
$\rho_*: \widetilde{K}_1(\Z[\pi]) \to \widetilde{K}_1({\rm \mathbb{M}}_n(\R))\cong \widetilde{K}_1(\R)
 \cong \R^+$.
Since $\R^+$ has no elements of finite order then there is the corresponding 
homomorphism ${\rm Wh}(\G) \to \R^+$. Therefore, an element $\omega \in {\rm Wh}(\G)$ 
has finite order if and only if $\rho_*(\omega)=1$ for every orthogonal representation 
$\rho$ of $\G$.

\paragraph{Elements of ${\rm Wh}(\G)$ as matrices and the representation dimension.} 
Nontrivial elements of the Whitehead group can be represented by matrices, usually
 of small size.
The {\it representation dimension} of a group $\G$ is said to be less than or equal to $m$,
with notation $r$-$\dim \G \leq m$, if every element of ${\rm Wh}(\G)$ can be realized as a matrix
in $GL(m, \Z[\G])$. 
If $\G$ is finite then $r$-$\dim \G \leq 2$. Furthermore, the representation dimension of the 
finite group $\G$ satisfies $r$-$\dim \G \leq 1$ if and only if
$\G$ admits no epimorphic mapping onto the following (see \cite{Sh})

\vspace{2mm}
\noindent  {\it 1.} the generalized quaternion group,

\noindent {\it 2.} the binary tetrahedral, octahedral, or icosahedral groups,

\noindent {\it 3.} and the groups 
$\Z_{p^2} \times \Z_{p^2}$, $\Z_p \times \Z_p \times \Z_p$,
$Z_p \times \Z_2 \times \Z_2 \times \Z_2$, $\Z_4 \times \Z_2 \times \Z_2$,
and $\Z_4 \times \Z_4$, for $p$ a prime.

\noindent Thus $r$-$\dim \G \leq 1$ for all finite simple groups. However, if we take 
products then the size of the matrix can grow (see expression \eqref{22} for an explicit matrix).



%

%
%
%



\section{
Further examples in heterotic M-theory}
\label{sec ex}
We have already seen many classes of examples both for the Whitehead group 
in section \ref{sec Wh} and for the Whitehead torsion in section \ref{sec tau}.
we now provide more examples and in particular ones which appear explicitly in 
model building (cf. the introduction). 

\paragraph{Tori and free abelian groups.} 
The fundamental group of the circle is the free abelian group $\Z$, so that the 
corresponding Whitehead torsion is zero,
${\rm Wh} (\Z)=0$.
For the $n$-torus $T^n$, the fundamental group
$\pi_1(T^n)=\Z^n$. This free abelian group of rank $n$ has a 
trivial  Whitehead torsion ${\rm Wh}(\pi_1(T^n))=0$, since
${\rm Wh}(\Z \oplus \cdots \oplus \Z)=0$ by the 
multiplicative property of Whitehead torsion under
free product (section \ref{sec Wh}).
It follows from the theorem of Bass 
about the rank of the Whitehead group that 
${\rm Wh}(\Gamma)$
of a free abelian group $\Gamma$ is zero if and only if $\Gamma$
has exponent 1, 2, 3, 4, or 6 \cite{Mil}.

\paragraph{Cyclic groups.}
Suppose $\G$ is a finite group. Then ${\rm Wh}(\G)$ is finitely 
generated, and ${\rm rank}({\rm Wh}(\G))$ is the difference between the number of irreducible
representations of $\G$ over $\R$ and the number of 
irreducible representations of $\G$ over $\Q$. 
For $\Gamma$ a cyclic group $\Z_p$ of order $p$, an odd prime, 
the numbers of representations are $q(\Z_p)=2$ and $r(\Z_p)=\frac{1}{2}(p+1)$, respectively.
This implies that 
${\rm Wh}(\Z_p)$ is the free abelian group of rank $(p-3)/2$ and that
${\rm Wh}(\Z_2)=0$. 
Alternatively, 
note that $\Z_p$ has $(p-1)/2$ inequivalent 
two-dimensional irreducible representations over $\R$, but one
$(p-1)$-dimensional irreducible representation over $\Q$ (since $\Q[\Z_p] \cong 
\Q \times \Q(\zeta)$, $\zeta$ a primitive $p$-th root of unity, and 
$[\Q(\zeta) : \Q]=p-1$), so 
${\rm rank}({\rm Wh}(\Z_p))=\frac{p-1}{2}+1 -2= (p-3)/2$. 
Note that we have already seen that ${\rm Wh}(\Z_k)=0$ for $k=2,3,4,6$.

\paragraph{Units in the group ring.}
Consider the integral group ring $\Z[\Z_p]$ of the finite cyclic group $\Z_p$ and let 
$\zeta$ be a primitive $p$th root of unity with corresponding group ring 
$\Z[\zeta]$. The pullback square of rings 
\(
\xymatrix{
\Z[\Z_p]
\ar[rr]
\ar[d]
&&
\Z[\zeta]
\ar[d]
\\
\Z 
\ar[rr]
&&
\mathbb{F}_p
}\;,
\)
where $\mathbb{F}_p$ is the field with $p$ elements, implies that the $(p-1)$st
power of any unit in $\Z[\zeta]$ comes from a unit in $\Z[\Z_p]$. An example of a unit in 
$\Z[\zeta]$ is $(\zeta + \zeta^{-1})^r$. This is invariant under complex conjugation 
in $\Z[\zeta]$ (this corresponds to invariance under the orientation duality 
discussed in section \ref{sec aut}).

\paragraph{The quintic and the cyclic group of order 5.}
The quintic threefold plays an important role as a prototype example
of compactification on Calabi-Yau manifolds. 
Consider the one-parameter family of quintic threefolds
$\cQ:= \{ z_1^5 + z_2^5 +z_3^5 + z_4^5 + z_5^5 + \psi^5 z_1z_2  z_3 z_4 z_5=0\}
\subset \C P^4$. The defining equation is invariant under 
the $\Z_5 \times \Z_5 \subset {\rm PGL}(5,\C)$ 
group action 
\(
[z_1 : z_2 : z_3 : z_4 : z_5]\mapsto [z_2 : z_3 : z_4 : z_5 : z_1]\;, \qquad
[z_1 : z_2 : z_3 : z_4 : z_5]\mapsto [\zeta z_1 : \zeta^2 z_2 :
\zeta^3 z_3 :
\zeta^4 z_4 :
 z_5]\;,
 \)
  where $\zeta=e^{2\pi i/5}$. The fixed points lie on $\CP^4-\cQ$, so that $\cQ/\Z_5$ and 
  $\cQ/\Z_5\times \Z_5$
are smooth Calabi-Yau threefolds. The six different $\Z_5$ subgroups in $\Z_5 \times \Z_5$
can be used. 
The Whitehead group of $\Z_5=\{t~|~t^5=1\}$ is ${\rm Wh}(\Z_5)=\Z$ with generator 
the torsion $\tau (u)$ of the unit $u=1-t+t^2 \in \Z[\Z_5]$ \cite{Mil}.
The identity $(t+t^{-1} -1)(t^2 + t^{-2} -1)=1$ indeed shows that $u$ is a unit. The 
homomorphism $\alpha: \Z[\Z_5] \to \C$, sending
$t$ to $\zeta$, also sends $\{\pm \g : \g\in \G\}$ to the roots of unity in 
$\C$, and 
hence $x \mapsto |\alpha (x)|$ defines a homomorphism from ${\rm Wh}(\Z_5)$ into 
$\R^*_+$, the nonzero positive real numbers. 
Then the map
$u \mapsto 1- \zeta - \zeta^{-1} =1- 2\cos(2\pi/5)$
can be used to show that no power of $u$ is equal to 1. 
Indeed, $|\alpha (u)|= | 1- 2\cos(2\pi/5)| \approx 0.4$, so that 
$\alpha$ defines an element of infinite order in ${\rm Wh}(\Z_5)$.
Note that the unit $u$ is self-conjugate, and that the automorphism 
$t \mapsto t^2$ of $\Z_5$ carries $u$ to $u^{-1}$. 
In fact, for $\Gamma$ finite abelian, every element of ${\rm Wh}(\Gamma)$ 
is self-conjugate \cite{Mil} (see the last paragraph in section \ref{sec tau}).

\vspace{3mm}
We see from the example of the quintic that, a priori,
there are countably infinitely many elements in the 
Whitehead group of the fundamental group of the quintic. 
Unless the Whitehead torsion  is the zero element,
there will be an obstruction to having a trivial $h$-cobordism 
and hence to a consistent relation to heterotic M-theory. Therefore, 
it is an interesting problem to compute the Whitehead 
torsion of the quintic.

\vspace{3mm}
Recall from the end of section \ref{sec Wh} that the full torsion subgroup of the
Whitehead group is given by $SK_1(\Z[\G])$. Therefore, one way to
tell that ${\rm Wh}(\G)$ is nontrivial is to detect torsion via $SK_1(\Z[\G])$.

\paragraph{Products of abelian groups.} 
We now consider products of abelian groups, in particular of cyclic groups.

\vspace{2mm}
\noindent {\it 1. Products of groups of even order.} For even order, 
we have already seen that the Whitehead group of the lowest rank 
non-simple group, $\Z_2 \times \Z_2$, is zero. Next we consider products of 
$\Z_2$ with $\Z_4$ and so on. We use the following two 
general formulae \cite{Ol} for the torsion part of the Whitehead group
$
SK_1(\Z[ (\Z_2)^k \times \Z_{2^n}])\cong
\left[ \oplus_{r=1}^k \binom{k}{r} \cdot (\Z_{2^{r-1}}) \right]
\oplus
\left[ \oplus_{s=2}^n (\Z_{2^s})\right]$
and $SK_1(\Z[(\Z_2)^2 \times \Z_{2^n}]) \cong \Z_2^{n-1}$.
For instance, 
the following cases can then be deduced: 

{\it 1.} $SK_1(\Z[\Z_4 \times \Z_4])\cong \Z_2$.

{\it 2.}  $SK_1(\Z[\Z_2 \times \Z_2 \times \Z_4])\cong \Z_2$.

{\it 3.} $SK_1(\Z[(\Z_2)^3 \times \Z_4])\cong (\Z_2)^3 \times \Z_4$.
This last  case is curious in that  ${\rm Wh}(\Gamma)=\Gamma$.

\vspace{3mm}
We can also use the general formula 
$SK_1(\Z[\Z_{4} \times \Z_{2^n}]) \cong (\Z_2)^{(n-1)}$
to deduce other relevant groups. For example, 
$SK_1(\Z[\Z_4 \times \Z_8])\cong (\Z_2)^2$,
$SK_1(\Z[\Z_4 \times \Z_{16}])\cong (\Z_2)^3$,
$SK_1(\Z[\Z_4 \times \Z_{32}])\cong (\Z_2)^4$, etc.

\vspace{2mm}
\noindent {\it 2. Products of groups of odd order.} Next we consider the case when 
the orders of the groups in the products are odd. 
We will look at groups of the form $(\Z_p)^k$, $\Z_{p^2} \times \Z_{p^n}$ and 
$(\Z_p)^2 \times \Z_{p^n}$, as well as combinations involving three factors, 
using general results from reference \cite{Ol}.

\noindent $(i)$ 
The torsion subgroup 
$SK_1(\Z[\G])$ is trivial if $\G$ is cyclic or an elementary 2-group, or 
of type $\Z_p \oplus \Z_{p^n}$. However, $SK_1(\Z[\G])$ is nontrivial 
form most abelian groups \cite{ADS}. 
If $\G=(\Z_p)^k$, $p$ odd, then $SK_1(\Z[\G])$ is a 
$\Z_p$-vector space of dimension 
$(p^k-1)/(p-1) - \binom{p+k+1}{p}$. 
For example, for $\G=(\Z_3)^3$, the torsion subgroup is 
$SK_1(\Z[(\Z_3)^3]) \cong (\Z_3)^3$.


\noindent $(ii)$ 
For $p$ an odd prime, $SK_1(\Z[\Z_{p^2} 
\times \Z_{p^n}]) \cong (\Z/p)^{(p-1)(n-1)}$.

\noindent $(iii)$ For $p$ an odd prime, $SK_1(\Z[(\Z_p)^2\times \Z_{p^n} ]) \cong 
(\Z_p)^{np(p-1)/2} 
$.

\noindent Let $p$ be an odd prime and $\Gamma$ an elementary abelian 
$p$-group of rank $k$. Then \cite{Ste} $SK_1(\Z[\Gamma])$ is an elementary
abelian $p$-group of rank 
$(p^k-1)/(p-1) - \binom{p+k-1}{p}$. In particular $SK_1(\Z[\Gamma])\neq 0$ for 
$k\geq 3$. For example, the following table can be formed (see also \cite{Ste})
\(
\begin{array}{|c|c|}
\hline
\Gamma & SK_1(\Z[\Gamma])\\
\hline
\hline
\Z_{p^2}\times \Z_{p^2} ~(p=3,5,7) & (\Z_p)^{p-1}\\
\hline
\Z_{p^2}\times \Z_{p}\times \Z_p ~(p=3,5,7) & (\Z_p)^{p(p-1)}\\
\hline
\Z_{27} \times \Z_9 & (\Z_3)^4\\
\hline
\Z_{27}\times \Z_3 \times \Z_3 & (\Z_3)^9\\
\hline
\Z_9 \times \Z_9 \times \Z_3 & (\Z_3)^{15} \times (\Z_9)^2\\
\hline
\end{array}
\)

\vspace{3mm}
\paragraph{Nonabelian groups.} We have already seen examples of nonabelian 
groups in section \ref{sec Wh}. In addition, 

\noindent {\it 1. Crystallographic groups.} $SK_1(\Z[\Gamma])=0$ for $\Gamma$
a dihedral, the binary tetrahedral or icosahedral group \cite{Mag} \cite{Ste}.

\noindent {\it 2. The quaternion group.} The Whitehead group ${\rm Wh}(\Z[Q_8])$ 
of the quaternion group $Q_8$ of order 8 is isomorphic to $\pm V$,
where $V=\Z_2\times \Z_2$ is Klein's 4-group \cite{Ke}. 
Note that $V$ is the factor group $Q_8/\{\pm\}$, where
$\{\pm\}$ is the commutator subgroup of $Q_8$.

\noindent {\it 3. Products with abelian groups.} If $\Gamma$ is any (nonabelian) 
quaternion or semidihedral 2-group, then 
for all $k \geq 0$, the torsion subgroup is
$
SK_1(\Z[\Gamma \times (\Z_2)^k]) \cong
(\Z_2)^{2^k-k-1}$.

\noindent {\it 4. Nonabelian groups with specified abelianization.}
For instance, if for order $|\Gamma|=16$ the torsion subgroup is given by 
\(
SK_1(\Z[\Gamma]) \cong 
\left\{
\begin{array}{ll}
1 & {\rm if~} \Gamma^{\rm ab} \cong \Z_2 \times \Z_2 {\rm ~~or~~} \Z_2 \times \Z_2 \times \Z_2\\
\Z_2 & {\rm if~} \Gamma^{\rm ab} \cong \Z_4 \times \Z_2.
\end{array}
\right.
\)

\paragraph{Finding Whitehead groups via transfer.}
Looking at inclusions tells us about the corresponding Whitehead groups.
We will consider several situations.

\noindent {\it 1.} Consider the cyclic group $\Z_{2k+1}$ of order $2k+1$
as a subgroup of the cyclic group $\Z_{4k+2}$ of order 
$4k+2$. Then the transfer $i^*: {\rm Wh}(\Z_{4k+2}) \to 
{\rm Wh}(\Z_{2k+1})$, corresponding to $i: \Z_{2k+1} \hookrightarrow \Z_{4k+2}$,
is onto for all $k$ 
\cite{Kw}. 

\noindent {\it 2.} Now consider the inclusion $i: \Z_{2k} \hookrightarrow \Z_{2k}\oplus \Z_2$.
Then the transfer $i^*: {\rm Wh}(\Z_{2k} \oplus Z_2) \to {\rm Wh}(\Z_{2k})$
is onto if and only if $k=1,2$ or 3 \cite{Kw2}. Since
${\rm Wh}(\Z_{2k})=0$ for $k=1,2$ and 3, then this means that 
${\rm Wh}(\Z_{2k} \oplus \Z_2)$ is trivial for these values of $k$. 

\noindent {\it 3.} Now let $\G$ be a finite abelian group of odd order. Then 
$i^*: {\rm Wh}(\G \oplus \Z_2) \to {\rm Wh}(\G)$ is onto \cite{Kw2}. 
This then can tell us whether $\G \oplus \Z_2$ is trivial from whether or not 
the Whitehead group of $\G$ itself is trivial. 

\vspace{3mm}
\noindent In general, if $\G \to \G'$ is a surjection of finite abelian groups 
induces a surjection $SK_1(\Z[\G]) \to SK_1(\Z[\G'])$
\cite{ADS}.

\paragraph{Semidirect products.}
For finite $\G$, the torsion subgroup of the Whitehead group is trivial  
$SK_1(R[\G])=1$ for all rings of integers in number fields if and only if
$\G$ is a semidirect product of two cyclic groups of relatively 
prime orders \cite{ADOS}.
In general, we can determine the ranks of the (torsion-free part) of 
these groups using Bass' theorem.

\vspace{3mm} 
Given the above rules and results, it is a straightforward exercise  to find the 
Whitehead groups of the fundamental groups appearing in the literature of
model building (reviewed partially in the introduction). 
 This includes, for instance, the groups appearing in \cite{Br}.

\vspace{3mm}
\paragraph{Whitehead torsion.}
The approach in this paper can also guide us to  
anticipate conditions on cobordisms when 
constructing Calabi-Yau threefolds with fundamental 
groups of certain types. Recall that just because the Whitehead 
group is nontrivial does not mean that the particular element,
the Whitehead torsion, is a nontrivial element. That is, one 
still has to compute the Whitehead torsion (geometrically), which 
we do not do here.  
We consider examples where 
elements in the torsion 
subgroup of the Whitehead group 
can be explicitly characterized (see \cite{Ol}).

\noindent $(i)$ For $\Gamma=\Z_4 \times \Z_2 \times \Z_2=
\langle g \rangle \langle h_1 \rangle \langle h_2 \rangle$, the torsion subgroup
is $SK_1(\Z[\Gamma]) \cong \Z/2$, and the nontrivial element is
represented by the matrix
\(
\left[
\begin{array}{cc}
1+ 8(1-g^2)(1+h_1)(1+h_2)(1-g) &
-(1-g^2)(1+h_1)(1+h_2)(3+g)\\
&\\
-13(1-g^2)(1+h_1)(1+h_2)(3-g) &
1+ 8(1-g^2)(1+h_1)(1+h_2)(1+g)
\end{array}
\right]
\in {\rm GL}(2, \Z[\Gamma])\;.
\label{22}
\)
In this case, one would have to check for a given $h$-cobordism 
built out of $Y^{11}$ and $M^{10}$ whether the corresponding 
Whitehead torsion is the zero element or the nontrivial element
represented by matrix \eqref{22}.

\vspace{2mm}
\noindent $(ii)$ For $\Gamma=\Z_3 \times Q_8=\langle g \rangle \times \langle a, b \rangle$, 
where $Q_8$ is a quaternion group of order 8, the torsion subgroup is
$SK_1(\Z[\Gamma])\cong \Z/2$,
and the nontrivial element is represented by the unit
\(
1+ (2-g-g^2)(1-a^2)\left(
3g + a + 4g^2a+4(g^2-g)b + 8ab
\right)
\in (\Z[\Gamma])^*\;.
\)
Again, one would check the geometry to see which of the two elements one gets.

\vspace{3mm}
It would be very interesting to calculate the Whitehead torsion explicitly for interesting 
classes of non-simply connected Calabi-Yau manifolds. As far as we know, 
no such calculations exist. One approach could be to find an explicit Morse
function (which seems not easy).

\section{Dynamical aspects}
\label{sec dyn}

In this section we consider some dynamical aspects of heterotic M-theory
as they arise in connection to the Whitehead group and Whitehead torsion.
We consider the effect of diffeomorphisms as well as orientation characters in 
section \ref{sec aut} and then consider dynamical constraints on general 
compactifications in heterotic M-theory in section \ref{sec com}.

\subsection{Automorphisms}
\label{sec aut}

\paragraph{Diffeomorphism.}
We study the effect of diffeomorphisms on our cobordisms, 
starting with a visible sector $M_0^{10}$.
Two eleven-dimensional cobordisms 
$(Y^{11}; M_0^{10}, f_0, M_1^{10}, f_1)$
and
$(Y'^{11}; M_0^{10}, f'_0, M_1'^{10}, f'_1)$
over $M_0^{10}$ are diffeomorphic relative $M_0^{10}$ 
if there is an orientation preserving diffeomorphism 
$F: Y^{11} \to Y'^{11}$ such that $F \circ f_0=f'_0$.
Indeed in \cite{FM} the quantum integrand in the M-theory effective action is 
shown to be invariant under the group of Spin diffeomorphisms of $Y^{11}$
which act freely on the space of metrics. 
On the other hand,
the effective action of the heterotic string is invariant under diffeomorphisms 
$\varphi: M^{10} \to M^{10}$ which lift to the Spin bundle and to the $E_8$
vector bundles \cite{W-tool}. The global anomaly is absent for arbitrary choices of the 
Spin $M^{10}$  and the two $E_8$ vector bundles. 

\vspace{3mm}
In addition to the many examples that we have considered so far,
one might be able to generate others using diffeomorphism. In a sense, 
constructing manifolds with cobordisms for which the Whitehead torsion
is nontrivial  would
be easier than calculating the Whitehead torsion for a given fixed cobordism. 
The idea is to take a cobordism and and glue it to another after a 
`twist' via an automorphism, i.e. a diffeomorphism in our case. This may give
rise to a nonzero Whitehead torsion.
This requires the study of the mapping torus as is done with the global anomalies 
in the heterotic effective action, e.g. in \cite{W-tool}.

\paragraph{Scale and intervals.} 
In the discussion so far we have used unit intervals $[0,1]$ to 
characterize the cobordism. In the physical set-up of Horava-Witten 
\cite{HW1} \cite{HW2} we have a length scale imposed by the 
dynamics in the theory. In the above formulation, we can introduce
this length scale by simply replacing the unit interval by the 
interval $[0, L]$ or $[-L, L]$, with $L$ the (dynamical) 
length in the eleventh direction.

\paragraph{Manifolds with non-positive sectional curvature.} 
It is interesting to note that ${\rm Wh}(\Gamma)$
is trivial for $\Gamma$ the fundamental group of 
closed manifolds with all the sectional curvatures $\leq 0$ \cite{FJ}. 
Therefore, although not Calabi-Yau (see \cite{HLW} Theorem 2.3), 
such spaces are admissible for $s$-cobordism (see \cite{Rares}).

\paragraph{The Whitehead torsion relative to left vs. right boundary.}
We ask whether it makes a difference to take the Whitehead torsion 
relative to the left boundary vs. taking it relative to the right boundary. 
There is a duality theorem which relates the Whitehead torsion relative to 
one boundary to that of the second boundary \cite{Mil}. For any 
orientable $h$-cobordism $(Y^{11}, M^{10}, M'^{10})$ we have the 
relation between $\tau (Y^{11}, M^{10})$ and $\tau (Y^{11}, M^{10})$ as
\(
\tau (Y^{11}, M'^{10})= \overline{\tau} (Y^{11}, M^{10})\;,
\)
 where
 $\overline{\tau}$ is the conjugate of $\tau$, defined as follows.
If $a=\sum n_i \g_i$ is an element of $\Z[\G]$, with $n_i\in \Z, \g_i\in\G$, then
 the conjugate of $a$ is
 the element $\sum n_i \g_i^{-1}$. This conjugation operation is an anti-automorphism 
 of the group ring with corresponding automorphism on $GL(\Z[\G])$ given by sending each
 matrix to its conjugate transpose. Passing to the abelianized group $K_1(\Z[\G])$ 
 gives an automorphism and hence an automorphism also of the quotient ${\rm Wh}(\G)$. 
 We see that `reversing' the direction of the cobordism, that is taking $M'^{10}$ to $M^{10}$
 instead of going from $M^{10}$ to $M'^{10}$, will result only in a mild 
 modification in having to deal with the conjugate torsion.  For large classes of examples
 in which we are interested, there is even a simplification. 
 If $\G$ is finite abelian then every element $\omega$ of ${\rm Wh}(\G)$ is self-conjugate,
 $\omega= \overline{\omega}$. This in particular holds for the distinguished element, 
 the Whitehead torsion. Therefore, for finite abelian fundamental groups working with 
 the Whitehead torsion relative to $M^{10}$ is equivalent to working with the Whitehead torsion
relative to $M'^{10}$.

\paragraph{Remark on the $E_8$ gauge bundles.} 
General boundary conditions for M-theory on a manifold with boundary are considered 
in \cite{DFM} \cite{DMW-boundary}. 
The left and right
 boundaries in heterotic M-theory each
carries an $E_8$ bundle which, in the process of model building is desired
to be broken down to a realistic group. Each of the two bundles is characterized
with a degree four characteristic class, $a_L$ for left and $a_R$ for right. 
As explained in \cite{DFM}, when $a_L=a_R$ then the eleven-dimensional 
spacetime provides a homotopy of the left and right connections so that 
the $E_8$ bundles on the
boundaries   
necessarily have $a_L=a_R$, which is the case in the non-supersymmetric
model in \cite{FH}. 
However, in (the supersymmetric) Horava-Witten theory, 
$a_L + a_R=\frac{1}{2}p_1(Y^{11})$. In order to overcome 
this difficulty, the authors of \cite{DFM} give a parity-invariant formulation 
of the C-field in M-theory by passing from $Y^{11}$ to $Y^{11}_d$, the orientation 
double cover of $Y^{11}$, and defining the C-field to be a parity invariant 
$E_8$ cocycle on $Y^{11}_d$. This is done via a nontrivial deck transformation
$\sigma$ on $Y^{11}_d$, so that a parity-invariant $E_8$ cocycle is 
one for which the differential character corresponding to the C-field 
satisfies $\sigma^* ([\check{C}])=[\check{C}]^{\cal{P}}$, where the action of the parity
${\cal P}$ is
$[\check{C}]^{\cal P}= [\check{C}]^*$. While this solves the parity problem it 
uses boundary conditions which lead to a Bianchi identity for the C-field which is 
different from the one in \cite{HW2}. 
We should keep 
these subtleties in mind when dealing with bundles, which are always there (but we 
do not directly deal with them in this paper). Nevertheless, 
next we provide an explanation of this in our current context. 


\paragraph{Orientation characters and twisted group algebras of the fundamental
group.}
The orientation character $\omega(M_0^{10}): \pi_1(M_0^{10}) \to \Z_2=\{\pm 1\}$
sends a loop $\g: S^{1} \to M_0^{10}$ to 
$\omega(\g)=+1$ (respectively, -1) if $\g$ is orientation-preserving
(respectively, orientation-reversing). Thus, in the oriented case $\omega (\g)=+1$ for 
all $\g \in \G$, that is $\omega$ is trivial if and only if $M_0^{10}$ is orientable. 
This has the following effect on the integral group ring of the fundamental group.
The orientation character defines a twisted involution (an anti-automorphism)
on the group ring $\Z[\G]$ given by $a \mapsto \omega(a) a^{-1}$, 
i.e. $\pm a$ according to whether $a$ is orientation preserving or reversing. 
The resulting group ring is denoted $\Z[\G]^\omega$. 

\vspace{3mm}
Let us consider this in more detail. 
An involution on $\Z[\G]$ is a function $\Z[\G] \to \Z[\G]$, taking an element 
$a$ to an element $\overline{a}$ satisfying:
$\overline{(a + b)}=\overline{a} + \overline{b}$,
$\overline{(ab)}=\overline{b}\cdot \overline{a}$,
~$\overline{\overline{(a)}}=a$, and $\overline{1}=1\in \Z[\G]$. 
This gives rise to the $\omega$-twisted involution on $\Z[\G]$, defined as
the map from $\Z[\G]$ to $\Z[\G]$ given by 
\(
a=\sum_{\g \in \G} n_{\g}\g \longmapsto \overline{a}=\sum_{\g \in \G} \omega(\g) n_\g \g^{-1}\;,
\quad n_\g \in \Z.
\)
In this case we have to use $\omega$-twisted cohomology and fundamental class in evaluating
expressions in the theory. 
Starting from the cellular $\Z[\pi_1(M_0^{10})]$-module chain complex $C(\widetilde{M}_0^{10})$,
the $\omega(M_0^{10})$-twisted involution on $\Z[\pi_1(M_0^{10})]$ can be used to 
define the left $\Z[\pi_1(M_0^{10})]$-module structure on the dual cochain complex
$C(\widetilde{M}_0^{10})={\rm Hom}_{\Z[\pi_1(M_0^{10})]}\left( C(\widetilde{M}_0^{10},
\Z[\pi_1(M_0^{10})] \right)$.
When $M_0^{10}$ is compact,
\footnote{$M_0^{10}$ does not necessarily have to be a manifold, but just a Poincar\'e
duality complex.}
 the fundamental class
 is given by $[M_0^{10}] \in H_{10}(M_0^{10}; \Z^{\omega(M)})$ such that the cap product
defines $\Z[\pi_1]$-module isomorphisms 
\(
[M_0^{10}] \cap - ~: H^*_{\omega(M)} (\widetilde{M}_0^{10}) \buildrel{\cong}\over{\longrightarrow}
H_{10-*}(\widetilde{M}^{10}_0)
\)
with $\widetilde{M}^{10}_0$ the universal cover of $M_0^{10}$. 
Quantities, e.g. ones appearing the effective action and the 
corresponding partition function, should be formulated using this fundamental class.

\vspace{3mm}
There is a duality formula for the Whitehead torsion which takes into account the 
orientation character. Let $Y^{11}$ be an eleven-dimensional $h$-cobordism
and let $\omega: \Gamma \to \{\pm 1\}$ be the orientation character. This gives rise to 
an anti-involution on the integral group ring $\Z\G$ by sending a group element 
$g$ to $\omega g^{-1}$, as above, and hence leads
to an involution $*$ on the Whitehead group ${\rm Wh}(\G)$. Then 
Milnor's duality formula is cast as 
\(
\tau (Y^{11}, M'^{10})= \tau (Y^{11}, M^{10})^*\;.
\)

\paragraph{Effect on F-theory.} Recently there has been a lot of research activity in 
model building using F-theory (see \cite{De} and references therein). 
F-theory can be considered as a limit of M-theory on a 2-torus when the 
volume of the two-torus becomes very small. This means that constraints on
the possible fundamental groups of $Y^{11}$, assumed to have a 2-torus factor, 
will have an effect on the possible fundamental groups on the space on 
which F-theory is considered.
Nontrivial fundamental groups in this context are considered in \cite{Br2}.
Therefore, we expect that our discussion in the heterotic/M-theory setting will
have, via duality, consequences for fundamental groups in F-theory.
 This is strengthened by the fact that 
in a class of models which admit perturbative heterotic duals, the F-theory and heterotic
computations match \cite{DW}.
It would be interesting to perform explicit checks of this in relevant examples.

\subsection{Compactification}
\label{sec com}

We have considered in general the relation between M-theory on a general eleven-manifold
and heterotic string theory on a general ten-manifold $M_0^{10}$. 
There are two aspects to this. First, for consistency the theory should make sense on any admissible
manifold and so studying this might give insight into understanding the theory further.
Second, there are certain favorable types of spaces for model building. 
We have in mind that 
$M_0^{10}$ is a product (or a bundle) of a Calabi-Yau threefold $X^6$ with a 
four-dimensional spacetime. In general, the latter can be taken to be a general 
four-manifold that solves the equations of motion and does/does not break supersymmetry
according to the goal one has in mind. 
It can be taken to be flat Minkowski or something close. We study such situations in this section
and consider whether the choice of four-dimensional spacetime changes the discussion we
have had so far. 

\vspace{3mm}
We take M-theory on an eleven-manifold $Y^{11}=Z^7 \times N^4$, where 
$N^4$ is spacetime and $Z^7$ is a seven-dimensional cobordism of 
the Calabi-Yau threefold $X^6$. This always exists because the Stiefel-Whitney
numbers of a Calabi-Yau threefold are zero: $w_1=0$ because of orientation,
$w_2=0$ because of Spin, and $w_3=0$ because both $w_1$ and $w_2$ 
are zero; then the Stiefel-Whitney numbers $w_1w_5[X^6]$, 
$w_2w_4[X^6]$, and $w_3w_3[X^6]$ are all zero. The heterotic ten-manifold is 
of the form $M_0^{10}=X^6 \times N^4$.

\paragraph{The $h$-cobordism of a product.}
Let $(Z^7; X_0^6, X_1^6)$ be a seven-dimensional $h$-cobordism for the 
Calabi-Yau threefold $X_0^6$, and let $N^4$ be a closed four-manifold. 
Then we can form an eleven-dimensional 
$h$-cobordism $(Z^7 \times N^4; X_0^6 \times N^4, X_1^6\times N^4)$. 
From the cut and paste properties of the Whitehead torsion 
(see \cite{Mil} \cite{We} \cite{KL}), we get that the torsion are related as follows
\(
\tau (Z^7 \times N^4, X_0^6 \times N^4)= \tau (Z^7, X_0^6)~ \chi (N^4)\;,
\label{eq zcs}
\)
where $\chi(N^4)$ is the Euler characteristic of $N^4$. 
Thus the value of this invariant will determine whether there we can relate
the discussion of torsion in eleven/ten dimensions to that in seven/six
dimensions. The former is the global picture we have built so far, and the 
latter correspond to the actual situation studied in model building, that is 
the fundamental groups appearing as examples are those of $X^6$ and not
(necessarily) of $M^{10}$. 

\vspace{3mm}
If spacetime were compact and odd-dimensional then the Euler characteristic would vanish 
identically. In that case, the torsion would vanish. For example, if we take
spacetime to be the circle $S^1$ then 
$Z^7 \times S^1 \approx X_0^6 \times S^ \times [0,1] \approx X_1^6 \times S^1 \times [0,1]$, 
i.e. the torsion vanishes. In particular, this gives $X_0^6 \times S^1 \approx X_1^6 \times S^1$.

\paragraph{Product with a torus and Wall's finiteness obstruction.} 
The circle $S^1$ has fundamental group
$\pi_1(S^1)\cong \Z$. If we consider the product $S^1 \times Y$, then what is the
corresponding Whitehead group in terms of that of the factors? There is in 
fact a direct sum decomposition \cite{BHS}
$
{\rm Wh}(\Z \times \Gamma) \cong {\rm Wh}(\Gamma) \oplus \widetilde{K}_0(\Z[\Gamma])
\oplus N$,
for some Nil-group $N$. For the 2-torus with fundamental group $\Z^2$, the process can be repeated.
It might seem that for this product we can have nonzero Whitehead group for the product manifold 
even though that group for the factors might not be zero. However, elements in 
$\widetilde{K}_0(\Z[\pi_1(X)])$, called Wall's finiteness obstruction, detects whether or
not $X^6$ has the homotopy type of a CW-complex. If we are within the category of such 
spaces then this element within the class group vanishes.   

\paragraph{Spacetime with flat structure.}
A manifold admits a flat structure if the tangent bundle is isomorphic to a flat vector
bundle, i.e. admits a flat connection. Even for such manifolds, 
one can have nonzero Euler-characterstic.
 For example, if we take take the connected sum
 $N^4=(\Sigma_3 \times \Sigma_3)~ \#_{i=1}^6(S^1 \times S^3)$,
 where $\Sigma_3$ is a surface of genus 3. The product $\Sigma_3 \times \Sigma_3$ is 
almost parallelizable and the product of spheres $S^1 \times S^3$ is parallelizable. 
Then the Euler characteristic is $\chi(N^4)=4$ (see \cite{Sm}).  In this example, the fundamental group 
is the free product $\pi_1(N^4)=\G_1 \ast \G_2$, where $\G_1$ is the direct product of 
two copies of a non-abelian surface group and $\G_2$ is of rank 6. 
In fact, $S^1 \times S^3$ can be replaced by any parallelizable four-manifold.

\paragraph{Compact vs. noncompact spacetime.} 
So far we have taken $N^4$ to be compact.
For compact manifolds, 
the existence of a smooth Lorentzian metric is equivalent to the manifold
having a vanishing Euler characteristic (see \cite{St}).
However, the situation gets modified in the presence of singularities
(see \cite{Ma}). 
What if it is not compact? 
Noncompact spacetimes are more desirable for the purpose of 
equipping spacetime with a Lorentzian structure; 
all noncompact
manifolds admit a Lorentzian metric. 
On the other hand, every noncompact manifold admits vector fields
with any specified set of isolated zeros. This suggests that 
noncompact manifolds with
nonzero (appropriate notion of)
\footnote{Note that there are various definitions and versions of the Euler characteristic in 
the noncompact setting.}
 Euler characteristic are abundant. 
Note that for noncompact Riemann surfaces, the Cohn-Vossen theorem
gives the inequality $\int_\Sigma K dA \leq 2\pi \chi(\Sigma)$
(see e.g. \cite{Ku}). In general one works with $L^2$-Euler characteristics. 
For example, the Euler characteristic of an Asymptotically Locally Euclidean 
(ALE) space corresponding to the Lie algebra of type $A_n$ is $n+1$.  
It is important to note that it should be checked whether equation \eqref{eq zcs} 
extends to the noncompact case. 
Furthermore, strictly speaking, in the noncompact case we have to use 
the noncompact version of the $s$-cobordism theorem,
for which the Whitehead torsion lives a new group, which fits into an 
exact sequence involving the Whitehead group and algebraic $K_0$, as well
as information about the ends \cite{Si2}.
Some aspects of behavior of ends in M-theory are discussed in \cite{DMW-corner}.

\vspace{3mm}
In the following few paragraphs we describe  a way for studying the 
Whitehead torsion via other invariants, namely the Reidemeister torsion \cite{Mil}
and the Ray-Singer torsion \cite{RS}. This then provides a setting for making some 
direct connections to phenomenology.

\paragraph{Relation of the Whitehead torsion to Reidemeister torsion.}
The Whitehead torsion $\tau$ is closely related to Reidemeister torsion (or R-torsion)
$\Delta$;
the former generalizes the latter but is a more delicate invariant. 
Algebraically, the Whitehead torsion is more general than R-torsion in that it is 
also defined for noncommutative rings (such as the group ring of the fundamental 
group $\Z[\pi_1(X)]$) for which the determinant, needed for the R-torsion, is 
not defined. The 
R-torsion is a 
topological invariant which distinguishes spaces which are 
homotopy equivalent but not homeomorphic, and is defined 
for spaces whose fundamental group $\pi$ is finite and for which 
the homology with coefficients in a certain $\pi$-representation vanishes. 
The R-torsion is defined in more general situations than Whitehead 
torsion, since any homotopy equivalence is a homology equivalence. 
Furthermore, R-torsion has two advantages over the Whitehead torsion:

(i) It is more likely to be defined.

(ii) Its value is an honest real number, instead of being an element of a
somewhat esoteric group. 

\noindent On the other hand, when defined, the Whitehead torsion is a sharper invariant. 
When they are both defined, the R-torsion is a function of the Whitehead torsion.
That is, for each unitary (orthogonal) representation
$\rho$  of the fundamental group $\pi$, the 
R-torsion is the real part of the determinant of the complex (real) matrix  
induced by $\rho$ from any matrix representation of the Whitehead torsion. 
One can find a useful criterion for when the Whitehead torsion is zero by
studying the R-torsion. 
For concreteness, let $h : \pi_1(M^{10}) \to O(n)$ be an orthogonal representation 
of the fundamental group $\pi=\pi_1(M^{10})$. Then $h$ extends to a unique homomorphism
from the group ring $\Z[\pi]$ to the ring ${\cal M}_n(\R)$ of all real $n \times n$ matrices
and determines a homomorphism $h_*: {\rm Wh}(\pi) \to \overline{K}_1(\R)
\cong \R^+$. Suppose that the Whitehead torsion $\tau(Y^{11}; M^{10}) \in {\rm Wh}(\pi)$
is defined and 
 suppose that $\pi$ is a finite group. Then it follows from the identity relating the 
 two torsions
 \cite{Mil}
\(
\Delta_h(Y^{11}; M^{10})= h_*\tau (Y^{11}; M^{10})
\)
that  $\tau(Y^{11}; M^{10})$
is an element of finite order in ${\rm Wh}(\pi)$ if and only if the R-torsion 
is $\Delta_h(Y^{11}; M^{10})=1$ for all possible orthogonal representations $h$ of 
$\pi$. If $\pi$ is finite abelian, then $\tau (Y^{11}; M^{10})=0$ if and only if 
$\Delta_h(Y^{11}; M^{10})=1$ for all possible such representations $h$. 
Since the R-torsion is easier to calculate, 
this gives a concrete way of checking whether the Whitehead torsion vanishes without
having to go through the difficult task of calculating it explicitly.

\paragraph{Examples of when R-torsion is defined and the Whitehead torsion is not.}
There are examples in which the Whitehead torsion cannot be defined 
but the R-torsion can (see \cite{Mil}). For instance, the Whitehead torsion $\tau (S^1)$ of the 
circle $S^1$ cannot be defined since the module $H_0(\hat{S}^1)$ for the 
universal cover $\hat{S}^1$ is not zero, and is not a free $\Z [\pi]$-module. 
On the other hand, the R-torsion is defined; if the homomorphism 
$h$ from the fundamental group $\pi_1(S^1)$ to the units $\mathbb{F}^\times$ in 
a field $\mathbb{F}$ maps a generator into the field element $x\neq 1$, then 
the associated R-torsion $\Delta_h(S^1) \in \mathbb{F}^\times /\pm h(\pi_1)$
is well-defined and equal to $1-h$, up to multiplication by $h^{m}$ for some $m\in \Z^\times$. 
Another example is a knot complement $X$ in the 3-sphere with $h: \pi_1(X) \to \mathbb{F}^\times$
mapping each loop with linking number $+1$ into the field element $x \neq 1$. 
Then the R-torsion is well-defined, and is equal to $(1-h)/A(h)$, where 
$A(h)$ is the Alexander polynomial of the knot. 

\paragraph{Effect on phenomenology.}
The Ray-Singer torsion, which is an analytic analog of R-torsion and which coincides
with it for Riemannian manifolds, has direct physical applications. 
The Ray-Singer torsion can be defined using determinants of Laplacians. 
In this form it has natural connection to one-loop amplitudes. For example, 
this torsion governs the threshold corrections for the heterotic string \cite{BCOF}.
In M-theory compactifications on manifolds with $G_2$ holonomy, the 
GUT scale $M_{\rm GUT}$ is essentially given by the Ray-Singer torsion 
$\Delta_{RS}(\Sigma)$ via $M_{\rm GUT}^3= \Delta_{RS}(\Sigma)/V_\Sigma$, where 
$V_\Sigma$ is the volume of the corresponding 3-cycle $\Sigma$ \cite{FW}.
For example, when $\Sigma=S^3/\Z_q$ is a lens space, on which there is a Wilson line 
of eigenvalues 
$\left( e^{2\pi i(2m/q)},  e^{2\pi i(2m/q)}, e^{2 \pi i(m/q)}, e^{-2\pi i(3m/q)}, 
e^{-2\pi i(3m/q)} \right)$ with $m$ and $q$ coprime integers, then 
the Ray-Singer torsion for the lens space is $\Delta_{RS}(\Sigma)=4q\sin^2(5\pi m/q)$.
Now, the more delicate Whitehead torsion can be partially studied by considering 
the R-torsion (or Ray-Singer torsion) as above. It should be an obstruction 
to supersymmetry in heterotic M-theory. The breaking scale would be the intermediate 
5-dimensional scale, and only gravitationally mediate to the visible sector. 
It would be interesting to see how this works explicitly.

\paragraph{Higher-dimensional compactifications.} 
If we take our eleven-manifold $Y^{11}$ to be a product of two manifolds, where
the internal manifold is of dimension lower than 6 then we can no longer 
apply the $s$-cobordism arguments we have been using. In particular, the
$s$-cobordism  theorem fails  in dimensions five  and it is an open problem 
in dimension four (see \cite{CS} \cite{Ch}). 
For example, there exists an $h$-cobordism 
$(W^5, T^4, T^4)$, where $T^4$ is the four-dimensional torus, 
for which there is no diffeomorphism from $W^5$ to $T^4 \times [0,1]$.
Since ${\rm Wh}(\pi_1(T^4))=0$, the $s$-cobordism indeed fails in
five dimensions. For topological spaces, the theorem
fails in both four and five dimensions \cite{Si}; 
one might say that we could apply the $s$-cobordism in this case to the 
spacetime part rather than the internal part, now that spacetime has
grown to admissible dimensions. This certainly can be done and will 
give consistency conditions depending on fundamental groups of spacetime 
(the arguments we have outlined will go through with the obvious changes).
However, we would then not be studying fundamental groups 
for purposes of particle physics but rather for purposes of cosmology.

\vspace{15mm}
{\bf \large Acknowledgement}

\vspace{2mm}
The author would like to thank Jonathan Rosenberg for useful discussions
on the Whitehead torsion and Kenji Fukaya and the referee 
for useful comments. He also acknowledges the hospitality of the 
Department of Physics and the Department of Mathematics at the National 
University of Singapore where part of this work was done.


\end{document}